\documentclass[prl,reprint,footinbib,superscriptaddress,showpacs]{revtex4-1}
\usepackage{color}
\usepackage{graphicx,overpic}
\usepackage{amsmath,amssymb}
\usepackage{pstricks}
\usepackage{epstopdf}
\usepackage[normalem]{ulem}
\usepackage{bbold,units}
\usepackage{enumitem}

\usepackage{mathtools}

\DeclareFontFamily{U}{cry}{\hyphenchar\font=-1}
\DeclareFontShape{U}{cry}{m}{n}{ <-> cryst}{}
\newcommand{\cry}[1]{{\usefont{U}{cry}{m}{n} \symbol{#1}}}
\renewcommand{\vec}[1]{\underline{#1}}

\begin{document} 

\title{Group Theory of Chiral Photonic Crystals with $4$-fold Symmetry: \\ Band Structure and S-Parameters of Eight-Fold Intergrown Gyroid Nets}

\date{\today}

\author{Matthias Saba}
\email[E-mail:]{Matthias.Saba@fau.de}
\affiliation{Theoretische~Physik, Friedrich-Alexander~Universit\"at~Erlangen-N\"urnberg, 91058~Erlangen, Germany}
\author{Mark D.~Turner}
\affiliation{CUDOS~\&~Centre~for~Micro-Photonics, Swinburne~University~of~Technology, Victoria~3122, Australia}
\author{Klaus Mecke}
\affiliation{Theoretische~Physik, Friedrich-Alexander~Universit\"at~Erlangen-N\"urnberg, 91058~Erlangen, Germany}
\author{Min Gu}
\affiliation{CUDOS~\&~Centre~for~Micro-Photonics, Swinburne~University~of~Technology, Victoria~3122, Australia}
\author{Gerd E.~Schr\"oder-Turk}
\email[E-mail:]{Gerd.Schroeder-Turk@fau.de}
\affiliation{Theoretische~Physik, Friedrich-Alexander~Universit\"at~Erlangen-N\"urnberg, 91058~Erlangen, Germany}


\begin{abstract}
The Single Gyroid, or {\bf srs}, nanostructure has attracted interest as a circular-polarisation sensitive photonic material. We develop a group theoretical and scattering matrix method, applicable to any photonic crystal with symmetry $I432$, to demonstrate the remarkable chiral-optical properties of a generalised structure called {\bf 8-srs}, obtained by intergrowth of eight equal-handed {\bf srs} nets. Exploiting the presence of four-fold rotations, Bloch modes corresponding to the irreducible representations $E_-$ and $E_+$ are identified as the sole and non-interacting transmission channels for right- and left-circularly polarised light, respectively. For plane waves incident on a finite slab of the {\bf 8-srs}, the reflection rates for both circular polarisations are identical for all frequencies and transmission rates are identical up to a critical frequency below which scattering in the far field is restricted to zero grating order. Simulations show the optical activity of the lossless dielectric {\bf 8-srs} to be large, comparable to metallic metamaterials, demonstrating its potential as a nanofabricated photonic material. 
\end{abstract}

\maketitle

Dielectric photonic crystals (PC) and metallic metamaterials with chiral nanostructures attract interest because of their chiral-optical behaviour, including circular dichroism \cite{DeckerKleinWegenerLinden:2007}, negative refractive index \cite{Pendry:2004,ZhangParkLiLuZhangZhang:2009,*PlumZhouDongFedotovKoschnySoukoulisZheludev:2009} and optically-induced torque \cite{LiuZentgrafLiuBartalZhang:2010}. A particularly intricate design, inspired by its occurrence in butterfly wing scales \cite{MichielsenStavenga:2008,*SaranathanOsujiMochrieNohNarayananSandyDufresnePrum:2010,*SchroederTurkWickhamAverdunkBrinkFitzGeraldPoladianLargeHyde:2011}, is the {\em single Gyroid} (SG) or {\bf srs} net \cite{HydeOKeeffeProserpio:2008}. It forms in inorganic materials on various length scales \cite{MilleTyrodeCorkery:2013,VignoliniYufaCunhaGuldinRushkinStefikHurWiesnerBaumbergSteiner:2011,TurnerSchroederTurkGu:2011}, with several applications \cite{LuFuJoannopoulosSoljacic:2013,TurnerSabaZhangCummingSchroederTurkGu:2013,HurFrancescatoGianniniMaierHennigWiesner:2011,VignoliniYufaCunhaGuldinRushkinStefikHurWiesnerBaumbergSteiner:2011}. The prediction of circular dichroism for a {\bf srs} PC \cite{SabaThielTurnerHydeGuBrauckmannNeshevMeckeSchroederTurk:2011} has been experimentally verified \cite{TurnerSchroederTurkGu:2011,TurnerSabaZhangCummingSchroederTurkGu:2013}. The circular polarization discrimination observed in metallic {\bf srs} nets \cite{HurFrancescatoGianniniMaierHennigWiesner:2011} is lower than expected from the helical nature \cite{OhDemetriadouWuestnerHess:2013}. 

\begin{figure}[t]
        \hspace*{0.02\columnwidth}
	\begin{minipage}{.43\columnwidth}
	    \includegraphics[width=\textwidth]{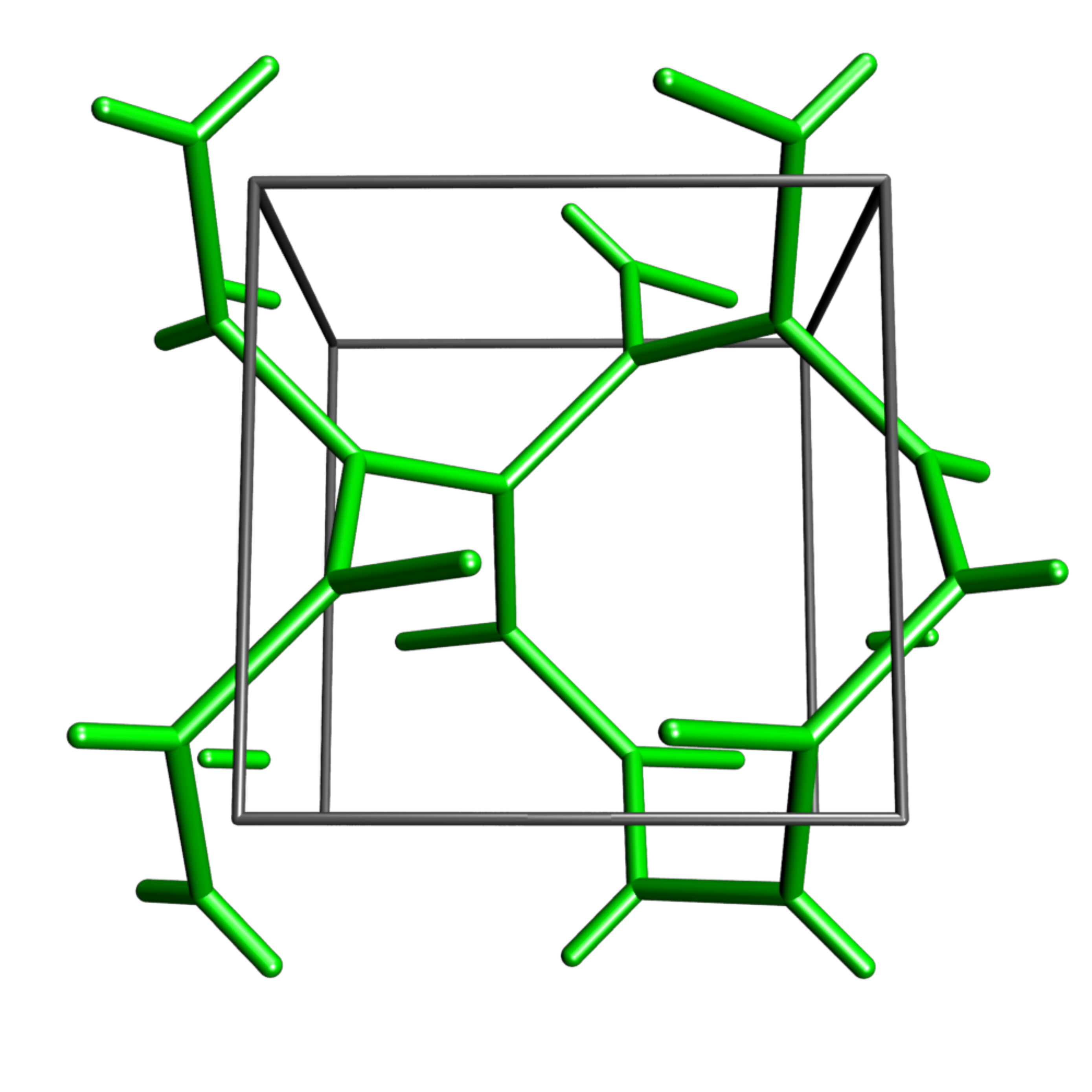}\vspace{-0.3cm}
	    {\bf 1-srs}: cubic $I4_132$ (214)
	\end{minipage} \hfill
	\begin{minipage}{.43\columnwidth}
	    \includegraphics[width=\textwidth]{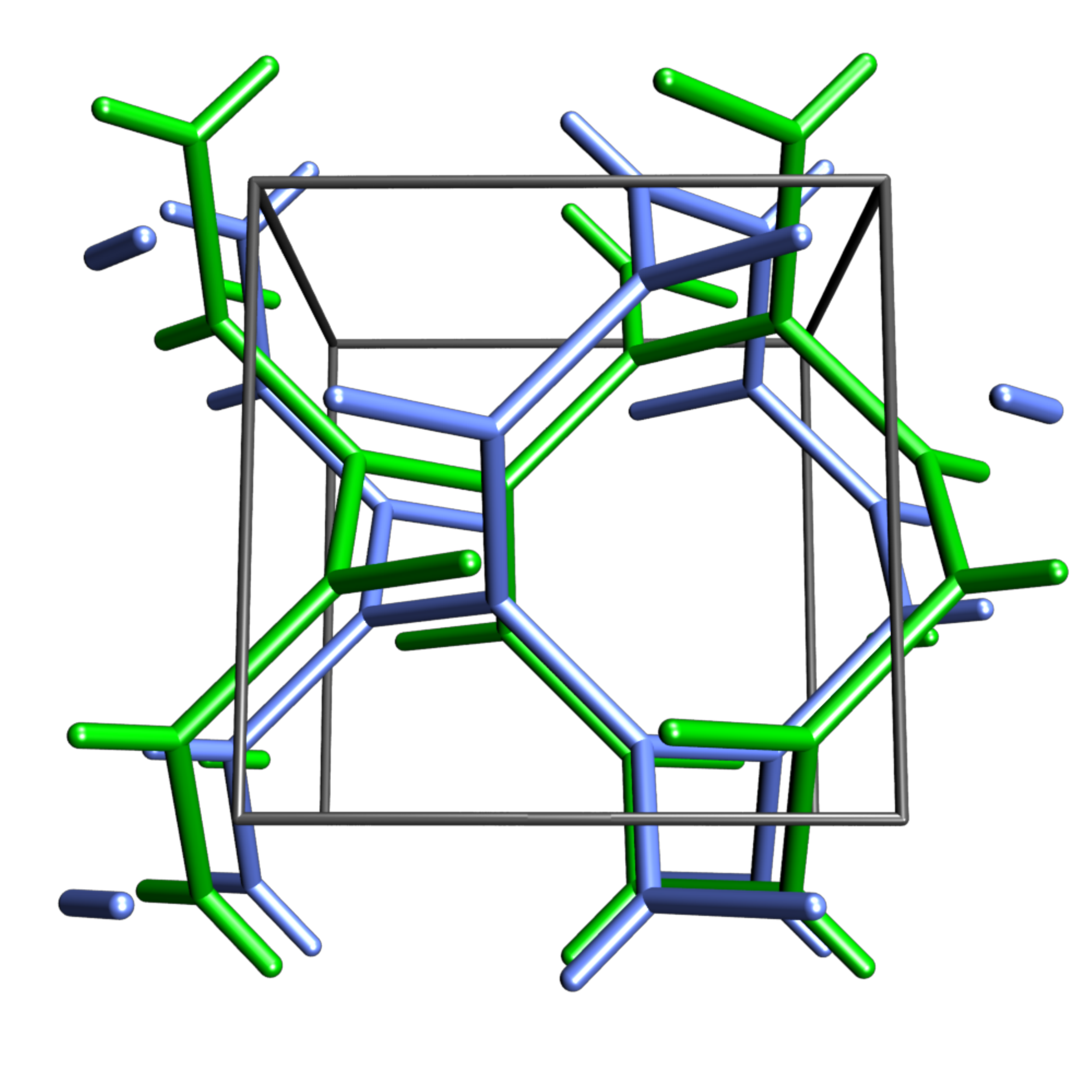}\vspace{-0.3cm}
	    \mbox{{\bf 2-srs}: tetragonal $P4_{2}22$ (93)}
	\end{minipage} 
        \hspace*{0.02\columnwidth}
        \\[0.2cm]
        \hspace*{0.02\columnwidth}
	\begin{minipage}{.43\columnwidth}
	    \includegraphics[width=\textwidth]{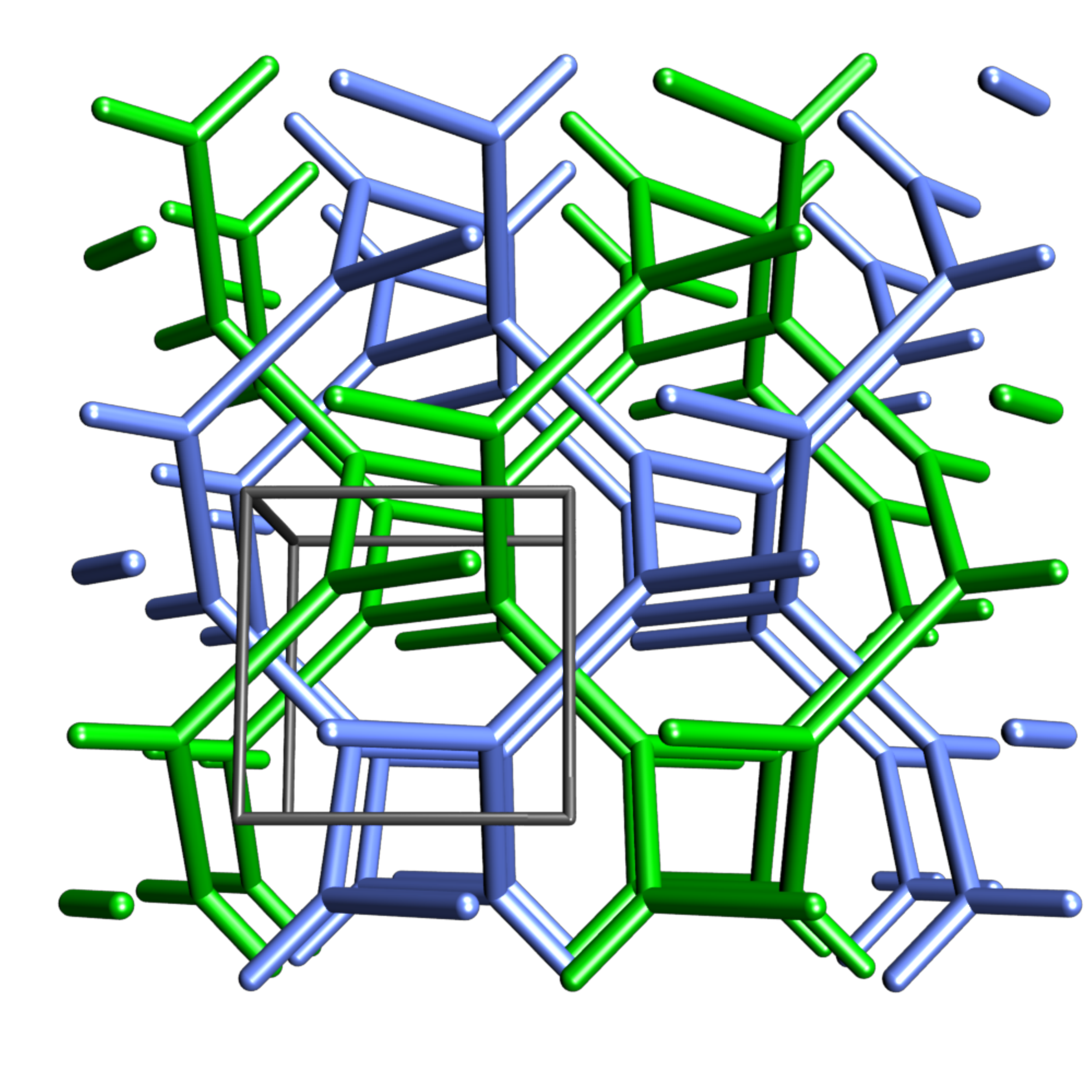}\vspace{-0.3cm}
	    {\bf 4-srs}: cubic $P4_232$ (208)
	\end{minipage} \hfill
	\begin{minipage}{.43\columnwidth}
	    \includegraphics[width=\textwidth]{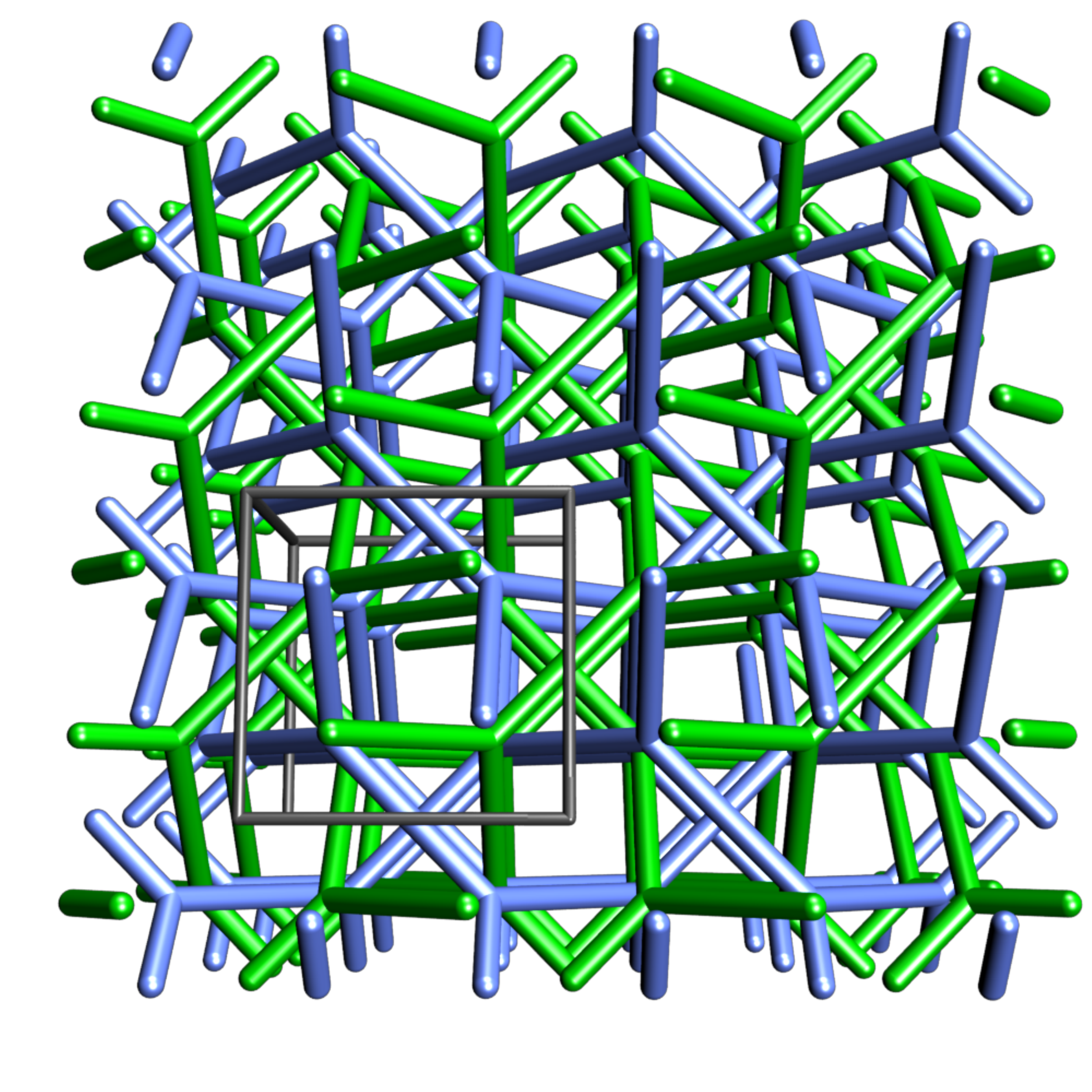}\vspace{-0.3cm}
	    {\bf 8-srs}: cubic $I432$ (211)
	\end{minipage}         
        \hspace*{0.02\columnwidth}
        \linebreak
	\caption{(Color Online) Construction of the {\bf 8-srs} by three replication steps. In each step the number of {\bf srs} nets is doubled by generating translated copies (blue) of the already existing nets (green). All nets are identical and equal-handed. }
\label{fig:1srs-8srs}
\end{figure}

Circular birefringence or optical activity (OA) and circular dichroism (CD) are polarisation effects related to the chiral properties of a light-transmitting medium. In the literature, OA and CD correspond to the difference in the absorption coefficients and refractive indices between left- (LCP) and right-circularly polarized (RCP) light of a homogeneous non-transparent material \cite{Hecht:1998}. Here, we adopt these terms for a slab of a lossless and inhomogeneous material, and relate OA and CD to the transmission and reflection amplitudes $(t,r)$. We define OA as the phase difference and CD as the relative difference in absolute values between the complex scattering amplitudes $s_\pm = t_\pm$ or $s_\pm = r_\pm$, respectively, for a respective incoming LCP ($+$) or RCP ($-$) plane wave:

\begin{equation}
	\text{CD}_s = \frac{\left|s_+\right| - \left|s_-\right|}{\left|s_+\right| + \left|s_-\right|}\text{;} \quad
	\text{OA}_s = \frac{\varphi^{(s)}_+ - \varphi^{(s)}_-}{2}\text{;} \quad e^{\imath\varphi^{(s)}_\pm} = \frac{s_\pm}{\left|s_\pm\right|}
        \nonumber
\end{equation}

While theoretical conclusions of this article, based on group theory and scattering matrix treatment, are valid for any structure with $I432$ symmetry (all nomenclature for symmetry groups as in \cite{InternationalTablesForCrystallography:1992}), we use a particularly interesting geometry called {\bf 8-srs} for illustration. The {\bf 8-srs} is a periodic structure consisting of eight identical, and hence equal-handed, inter-threaded copies of the {\bf srs} net \cite{HydeRamsden:2000}. 

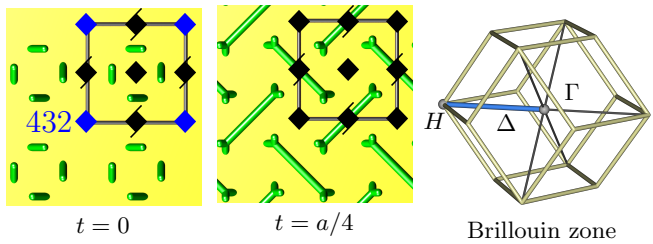
\begin{figure}[t]
\begin{minipage}{.3\columnwidth}
    \begin{overpic}[width=\textwidth]{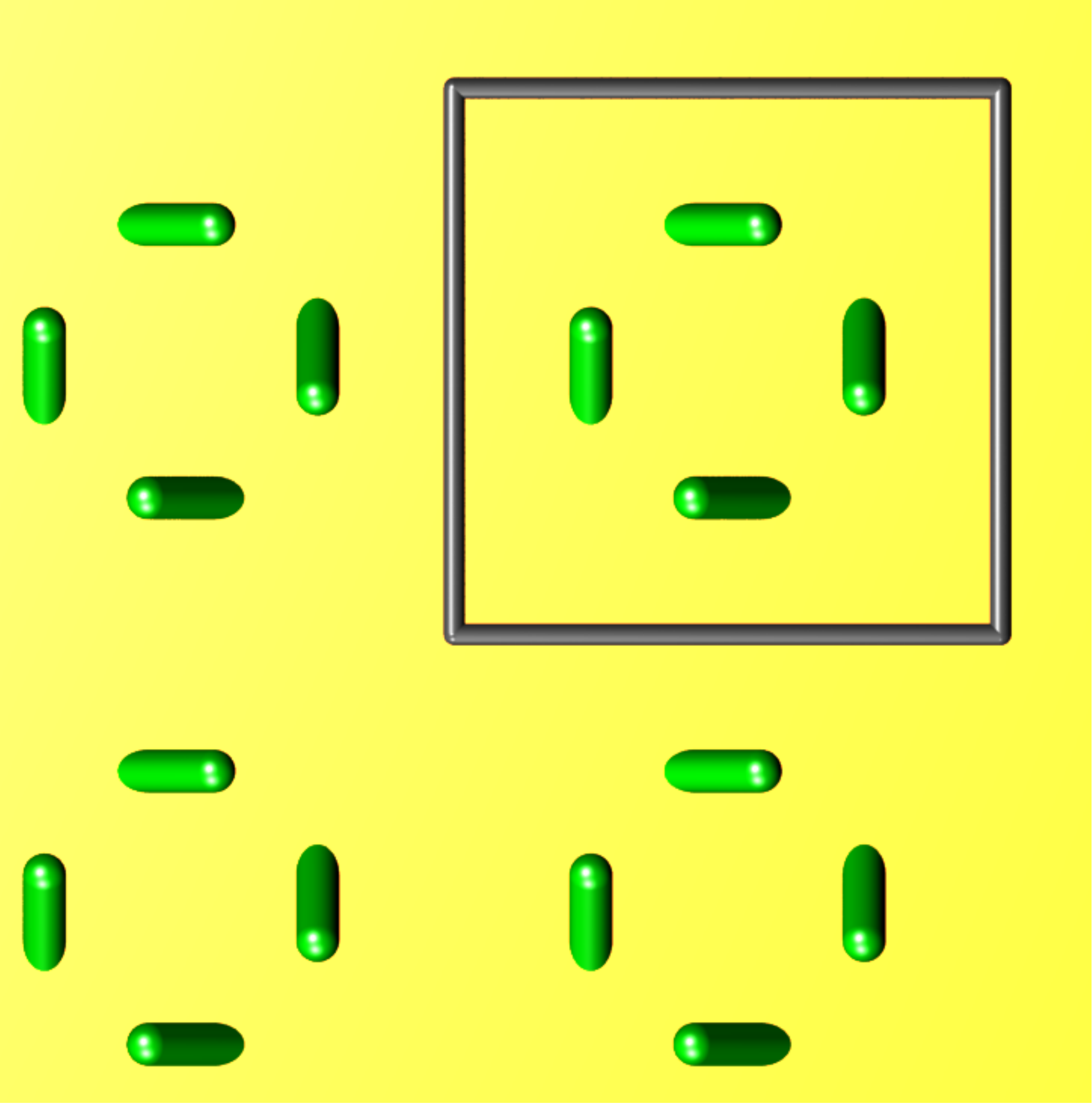}
    	\put(34.5,37.5){\color{blue} \Large \cry{4}}
    	\put(10,37.5){\color{blue} \large $432$}
    	\put(84.5,37.5){\color{blue}\Large \cry{4}}
    	\put(34.5,86.5){\color{blue}\Large \cry{4}}
    	\put(84.5,86.5){\color{blue}\Large \cry{4}}
    	\put(59.5,62){\Large \cry{4}}
    	\put(59.5,37.5){\Large \cry{42}}
    	\put(59.5,86.5){\Large \cry{42}}
    	\put(34.5,62){\Large \cry{42}}
    	\put(84.5,62){\Large \cry{42}}
    \end{overpic}\\
    \centering
    $t=0$
\end{minipage} \hfill
\begin{minipage}{.3\columnwidth}
    \begin{overpic}[width=\textwidth]{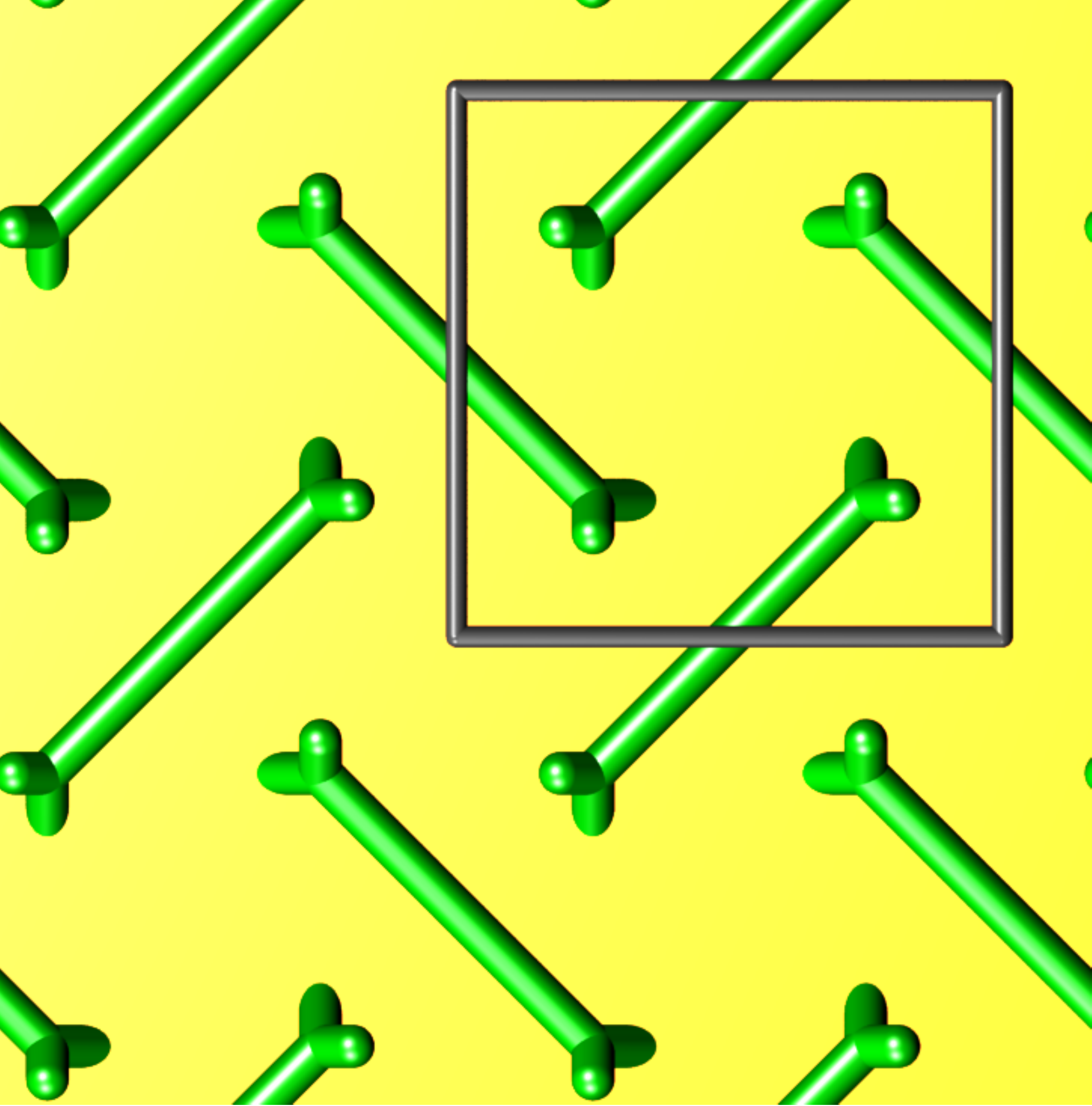}
    	\put(35,37.5){\Large \cry{4}}
    	\put(85,37.5){\Large \cry{4}}
    	\put(35,86.5){\Large \cry{4}}
    	\put(85,86.5){\Large \cry{4}}
    	\put(60,62){\Large \cry{4}}
    	\put(60,37.5){\Large \cry{42}}
    	\put(60,86.5){\Large \cry{42}}
    	\put(35,62){\Large \cry{42}}
    	\put(85,62){\Large \cry{42}}
    \end{overpic}\\
    \centering
    $t=a/4$
\end{minipage} \hfill
\begin{minipage}{.35\columnwidth}
	\begin{overpic}[width=\textwidth]{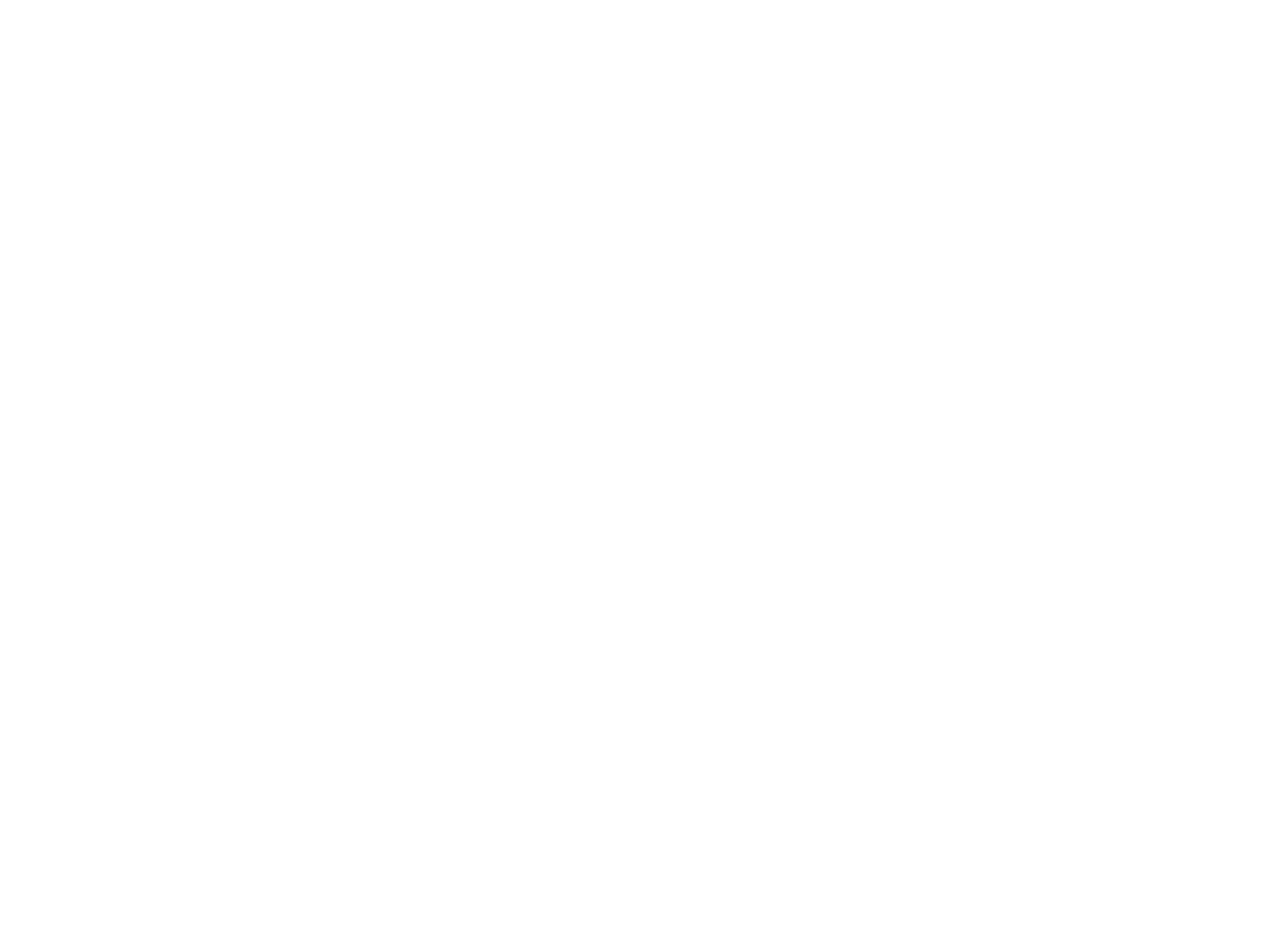}
		\put(0,-25){\includegraphics[width=1\textwidth]{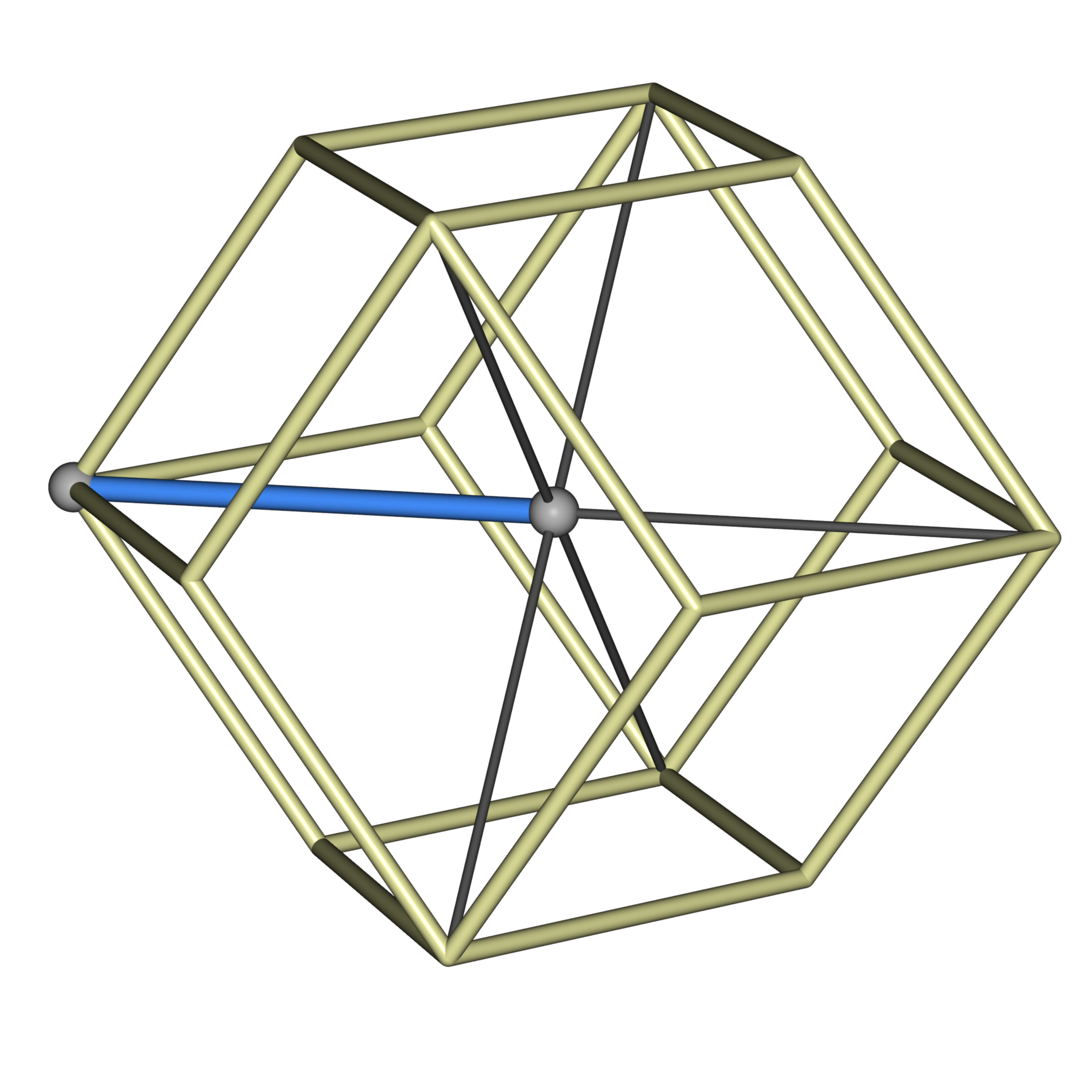}}
		\put(-2,19){$H$}
		\put(30,18){$\Delta$}
		\put(60,32){$\Gamma$}
	\end{overpic}\vspace{.5cm}
	\centering
	Brillouin zone
\end{minipage}
\caption{(Color Online) Left and center: Different cross-sections with $[100]$ inclination through the {\bf 8-srs} reveal its four-fold symmetry.  $4$-fold rotation and $4_2$ screw axes are marked by the symbols \cry{4} and \cry{42}, respectively. The grey square represents the cross section of the cubic unit cell with vertices at the $432$ (Schoenfliess $O$) symmetry point (\cry{4}). Right: The BCC Brillouin zone is a rhombic dodecahedron. The crystallographic $[100]$ directions are marked by thin black lines. The blue $\Delta$ line connects the high symmetry points $\Gamma$ and $H$.}
\label{fig:8srs-cross-section}
\end{figure}

Figure \ref{fig:1srs-8srs} shows that the {\bf 8-srs} is obtained by arranging translated copies of the {\bf srs} net, such that all eight networks remain disjoint and to give body centered cubic symmetry $I432$ \footnote{See also the reticular chemistry structure resource {\tt www.rscr.anu.edu.au} \cite{OKeeffePeskovRamsdenYaghi:2008} for details, where the {\bf 2-srs}, the {\bf 4-srs} and the {\bf 8-srs} are denoted srs-c2*, srs-c4 and srs-c8, respectively.}. With $a_0$ the lattice parameter of the {\bf 1-srs} in its symmetry group $I4_132$, adding a copy translated by $a\coloneqq a_0/2$ along $[100]$ gives the {\bf 2-srs} of tetragonal symmetry; translation along a distinct coordinate axis by $a$ yields the {\bf 4-srs} with simple cubic (SC) symmetry; translation by $\nicefrac{\sqrt{3}a}{2}$ along $[111]$ the {\bf 8-srs} with body-centered cubic (BCC) symmetry. Note that the {\bf 4-srs} and the {\bf 8-srs} have the same lattice constant $a=a_0/2$. Importantly, the {\bf 8-srs} has both four-fold rotation and four-fold screw axes along its [100] direction, see Fig.~\ref{fig:8srs-cross-section}, in contrast to the {\bf 1-srs} with only screw-rotations.

We consider the {\bf 8-srs} as a dielectric PC obtained by inflating all edges of the {\bf 8-srs} to solid struts (rods) with permittivity $\epsilon$, embedded in vacuum. For the simulations, we use $\epsilon=5.76$, close to high-refractive index Chalcogenide glass at telecommunication wavelengths \cite{NicolettiZhouJiaVenturaDouglasLutherDaviesGu:2008} or $\mathrm{Ti}O_2$ at optical wavelengths \cite{MilleTyrodeCorkery:2013}; the solid volume fraction is $\phi\approx31.4\%$, corresponding to a rod diameter of $d\approx0.23 a$. Finite size effects are obtained for a slab of size $\infty \times \infty \times N_z a$ with $[100]$ inclination of the PC \footnote{The infinite size in $x$ and $y$ direction is achieved by use of periodic boundary conditions assuming a single lateral unit cell of the {\bf 8-srs}.}. All analyses are for wave vectors $\vec{k}$ along that axis and assuming termination planes perpendicular to $[100]$. The parameter $t$ denotes the position of the termination plane in the unit cell.

By Bloch's theorem, the {\em translational} symmetry, or periodicity, of a PC leads to a natural representation of the eigenmodes by a set of orthogonal basis modes, the Bloch modes. The Bloch wave vector $\vec{k}$ characterises the translational symmetry behaviour of a mode and the band structure is the dependence of the frequency $\omega$ on $\vec{k}$.
% The translational symmetry of the {\bf 8-srs} has a direct impact onto its photonic eigenmodes. A set of orthogonal basis modes is naturally obtained by Bloch's theorem. The band structure is the dependence of those modes as a function of their translational symmetry behaviour characterized by the Bloch wave vector $\vec{k}$. 

\begin{figure}[t]
  \centering
	\begin{overpic}[width=0.95\columnwidth]{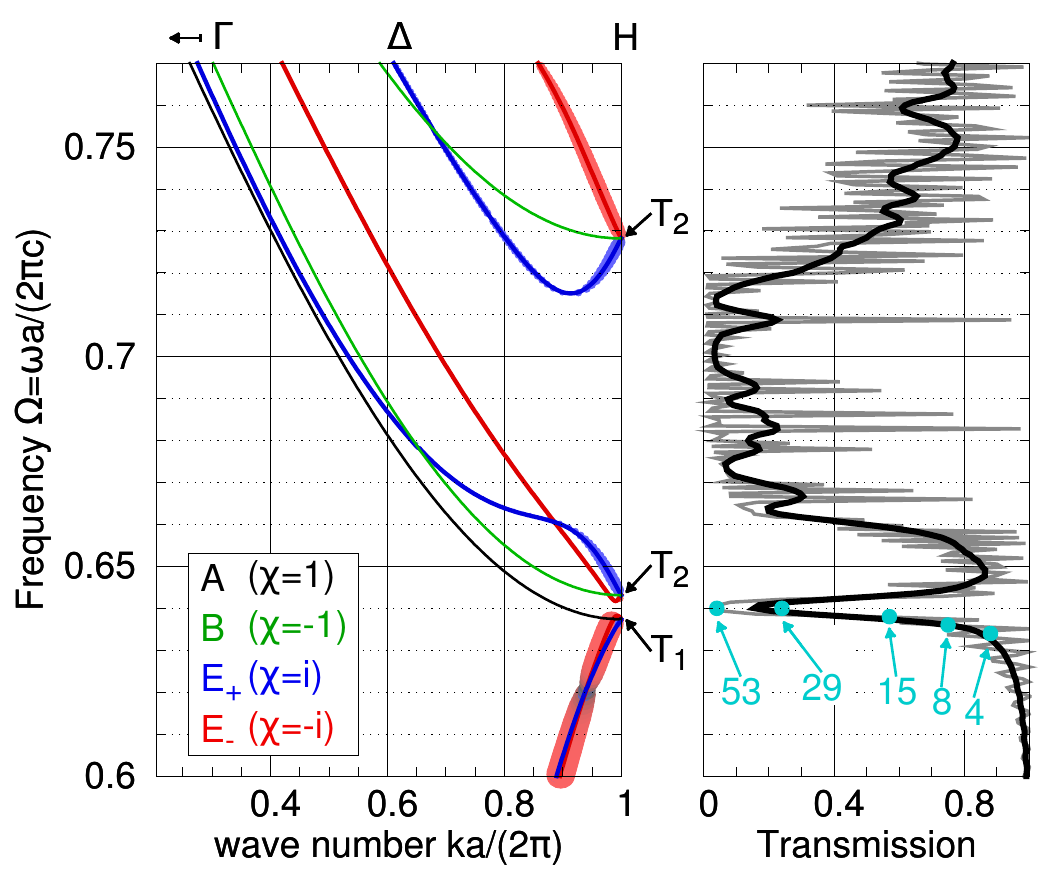}
	\end{overpic}
	\caption{(Color online) Band structure and transmission spectrum of the {\bf 8-srs} PC along $\Delta$ (Fig.~\ref{fig:8srs-cross-section}): (Left) Bands are colored according to mode's symmetry behaviour corresponding to the irreducible representations $i\in\{A,B,E_+,E_-\}$ of the $C_4$ point symmetry defined in Tab.~\ref{tab:O-C_4}. (Right) Transmission of light at normal incidence through slab of thickness $N_z=53$ and termination $t=0$. Transmission is the same for both LCP and RCP (thin light grey line). The black line is a convolution with a Gaussian with $\delta\Omega = 0.002$ that eliminates sharp Fano and Fabry-P\'erot resonances. Teal points mark the transmission minima at the \emph{pseudo}-bandgap at $\Omega\approx0.64$ for slabs of thickness $N_Z=4, \dots, 53$.}
	\label{fig:transmission-bandstructure}
\end{figure}

In analogy to the translational symmetries, we now classify the behaviour of the eigenmodes under point symmetries (here rotational symmetries). Group and representation theory provides the formalism for this classification, in the form of {\em irreducible representations} $A$, $B$, $E_\pm$ and $T_{1/2}$ and their {\em characters} that uniquely define these, see Tab.~\ref{tab:O-C_4}. We analytically obtain the following general relationships valid for any PC with $I432$ symmetry (including also the {\bf 8-srs}, see Fig.~\ref{fig:transmission-bandstructure}):\\[-0.8\baselineskip]
 
%Similarly, the point symmetries of the {\bf 8-srs} influence its eigenmodes. 
%We classify them by their point symmetry transformation behaviour with group theoretical analysis to obtain the following three general statements analytically (see Fig.~\ref{fig:transmission-bandstructure} for illustration)
%Numerical results are shown in Fig.~\ref{fig:transmission-bandstructure} and they are in perfect agreement with these general predictions.

\noindent {\bf (a) Three-fold degeneracy at the $H$-point:} The $4$ lowest eigenstates at the $H$ point are $3$-fold degenerate. There are two $T_1$ and two $T_2$ modes defined in Tab.~\ref{tab:O-C_4}) and classified by their respective point symmetry behaviour \footnote{The upper $T_1$ mode is out of frequency boundaries in Fig.~\ref{fig:transmission-bandstructure}.}.\\[-0.8\baselineskip]

\noindent {\bf (b) Degeneracy fully lifted on $\Delta$:} The degeneracy is fully lifted when going away from the high symmetry points onto the $\Delta$ line. Each mode split is summarized with a compatibility relation $T_1=A+E_++E_-$ or $T_2=B+E_++E_-$ (Tab.~\ref{tab:O-C_4}).\\[-0.8\baselineskip]

\noindent {\bf (c) Inversion symmetry and slope at $\Gamma$ and $H$:} Each band $\omega_i(k)$ along $\Delta$ is characterized by its {\it irreducible representation} $i\in\{A,B,E_+,E_-\}$. It has inversion symmetry $\omega_{\nicefrac{A}{B}}(-\vec{k})=\omega_{\nicefrac{A}{B}}(\vec{k})$ and $\omega_{E_\pm}(-\vec{k})=\omega_{E_\mp}(\vec{k})$. The bands $\omega_{\nicefrac{A}{B}}(\vec{k})$ hence approach the  points $T_1$ and $T_2$ with zero slope and the bands $\omega_{E_\pm}$ with equal and opposite slope.\\[-0.8\baselineskip]

\begin{figure}[t]
  \centering
	\includegraphics[width=0.95\columnwidth]{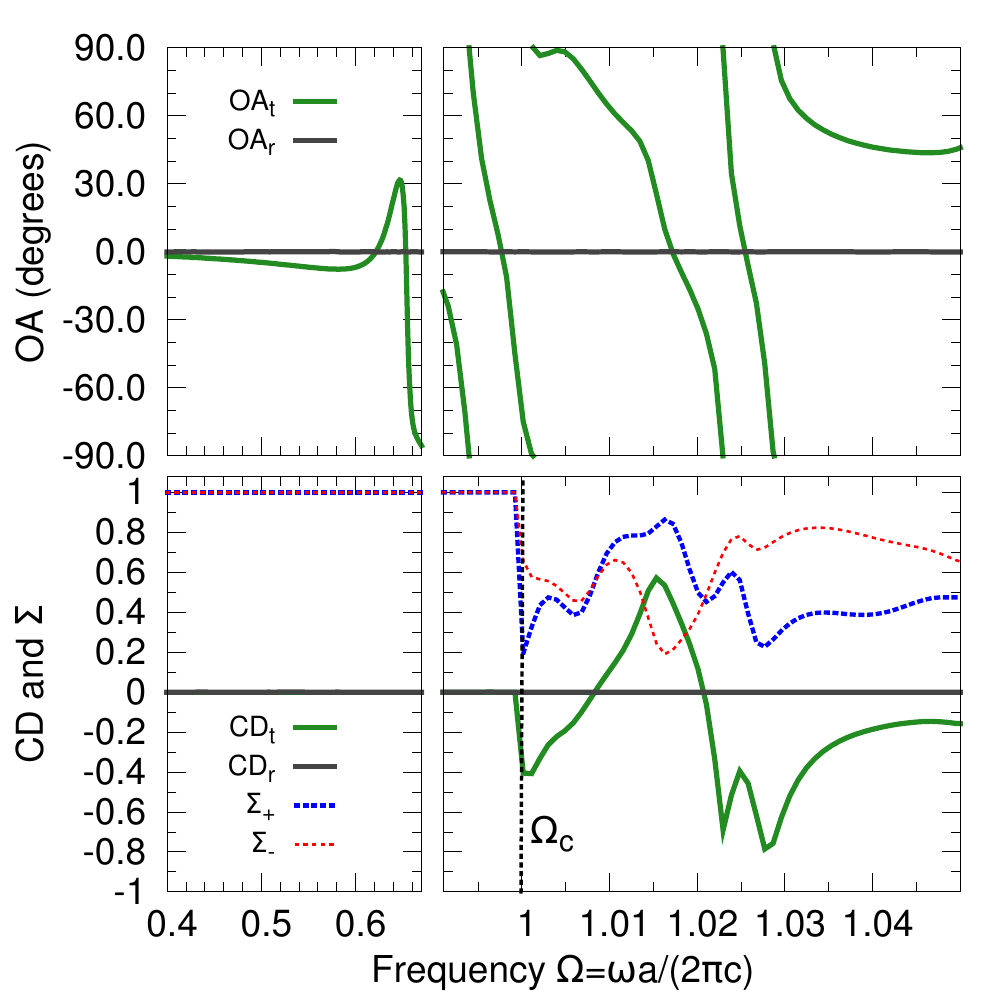}
        \caption{(Color online) Simulated CD, $\Sigma$ and OA for the reflection and transmission of a plane wave at normal incidence: The {\bf 8-srs} PC slab has termination $t=0.25a$ and thickness $N_z=4$. Since optical experiments cannot measure phase differences $> 90^\circ$, OA is wrapped onto the interval $[-90^\circ,90^\circ]$ by $\text{OA}\mapsto\arctan\left(\tan(\text{OA})\right)$.}
        \label{fig:optical-activity}
\end{figure}

Group theory combined with an analytic scattering matrix treatment yields the following four further general rules for photonic scattering at a finite slab of an $I432$ PC, inclined at $[100]$ direction and hit by a plane wave at normal incidence (see Fig.~\ref{fig:optical-activity} for simulations of the {\bf 8-srs}). These include previous results \cite{BaiSvirkoTurunenVallius:2007,MenzelRockstuhlLederer:2010,KaschkeGanselWegener:2012} as special cases.\\[-0.8\baselineskip]

\noindent {\bf (d) $\mathbf{\{A,B,E_+,E_-\}}$ correspond to non-interacting scattering channels, $E_-$ and $E_+$ represent RCP and LCP, respectively, A and B are dark modes:} Modes of distinct representation do not interact. Scattering takes place in four independent channels characterized by the four representations $A,B,E_+,E_-$. For each channel, a well-defined scattering matrix relating the amplitudes of the outgoing plane waves to those of the incoming plane waves is found. Both $A$ and $B$ representations represent {\it dark} modes that do not couple to any plane wave at normal incidence. At normal incidence, any $E_+$ mode couples only to RCP and any $E_-$ only to LCP plane waves. This implies that the channels corresponding to $A$ and $B$ do not contribute to the scattering process and there is no polarisation conversion between LCP and RCP in transmission and reflection at any wave length.\\[-0.8\baselineskip]

\noindent {\bf (e) No CD and OA in reflectance:} The reflection matrices on both sides are identical for the $E_+$ (LCP) and the $E_-$ (RCP) channel. For the reflection spectrum, CD and OA are hence strictly zero for all wave lengths.\\[-0.8\baselineskip]

\noindent {\bf (f) No CD in transmission below a critical frequency $\Omega_c$:} The matrix norm of the transmission matrices is identical for $E_+$ and $E_-$ channels. Henceforth, at low frequencies $\Omega\coloneqq\nicefrac{\omega a}{2\pi c}<\Omega_c\coloneqq1$, where the portion of energy that leaves the crystal in the $(00)$ Bragg order $\Sigma_\pm=|t_\pm|^2+|r_\pm|^2$ is strictly $100\%$, CD is zero. The matrix norm imposes no condition for circular dichroism above $\Omega_c$. Optical activity may be finite at any frequency.\\[-0.8\baselineskip]

The appendix outlines the proofs of the above claims. We now provide interpretations for the {\bf 8-srs} PC:

Due to rules (d)-(f), an {\bf 8-srs} slab of fixed thickness acts like an effective, optically active material for which the Kramers-Kronig relations are not valid; in contrast to homogeneous optically active materials, rotary power is not caused by a difference in the refractive indices for LCP and RCP but by the microstructure at the same length-scale as the wavelength of the light.

Rule (d) also provides an interpretation for our definition of OA and CD: If a linear polarised plane wave at normal incidence impinges on a finite slab, the perpendicularly scattered (zero Bragg order) wave is generally elliptically polarised. The principal axis of its polarisation ellipse is rotated by OA compared to the polarisation axis of the incoming wave and has eccentricity $e=\sqrt{1-(\text{CD})^2}$. For $\Omega < \Omega_c$, the polarization plane is hence rotated without introducing any ellipticity.

All results are accurately reproduced by numerical calculations. Fig.~\ref{fig:transmission-bandstructure} compares the transmission, which is the same for LCP and RCP light, through a thick slab with $N_z=53$ to the photonic band structure (PBS) of the infinite periodic PC. The PBS modes are colored according to their numerically determined irreducible representation \cite{SabaTurnerGuSchroeder:2013} and have a size proportional to the coupling constant $\beta$ \cite{SabaThielTurnerHydeGuBrauckmannNeshevMeckeSchroederTurk:2011} that describes the ability of the Bloch mode to couple to an incident plane wave of the same frequency \footnote{$\beta$ has been slightly improved compared to \cite{SabaThielTurnerHydeGuBrauckmannNeshevMeckeSchroederTurk:2011} by taking the field overlap integrals of the $4$ component vector $F=(H_\parallel,E_\parallel)^t$ instead of the three components of $H$ to take \emph{impedance mismatch} between the coupled fields into account. $\beta$ is still only a rough estimate for transmission especially when several (partly evanescent) Bloch modes are involved in the scattering process.}. The PBS shows the behaviour predicted by results (a)-(c): It is $3$-fold degenerate at the $H$ point and splits into $3$ separate bands of irreducible representation $A$ ($B$), $E_+$ and $E_-$. The slopes near the $H$ point are consistent with result (c). The PBS is further in good agreement with the transmission spectrum. Notably, the transmission drop at the frequency $\Omega_g\coloneqq0.64$ is fully consistent with our band structure results: Since the $A$ band has a dark mode behaviour according to rule (d), the band structure exhibits a small \emph{pseudo}-bandgap (with $T\rightarrow0$ for $N_z\rightarrow\infty$, cf.~teal points in Fig.~\ref{fig:transmission-bandstructure}) in the frequency range $0.637<\Omega<0.643$. Note that different choices of $\phi$ and $\epsilon$ give significantly larger bandgap width, see Fig.~$5$ in \cite{SabaTurnerGuSchroeder:2013}.

Fig.~\ref{fig:optical-activity} shows OA and CD spectra. In agreement with (e), $\text{OA}_r=\text{CD}_r=0$ to any numerical precision. Consistent with (f), $\text{CD}_t$ is only present above $\Omega_c$ where the $(10)$ Bragg order is non-evanescent. $\text{OA}_t$ is present at all frequencies and particularly strong above the fundamental bands with a large slope in the spectrum.
\begin{table}[t]
\begin{minipage}[t]{\columnwidth}
	\begin{minipage}[t]{0.58\textwidth}
		\begin{tabular}{|c|rrrrr|}
			\hline
			$O$ {\tiny $(C_4)$}		&	$\mathbb{1}$	&	$6C_4$	&	$3C_2$	&	$8C_3$	&	$6C_2^{'}$	\\
			\hline
			$A_1${\tiny $(A)$}		&	$1$		&	$1$	&	$1$	&	$1$	&	$1$		\\
			$A_2${\tiny $(B)$}		&	$1$		&	$-1$	&	$1$	&	$1$	&	$-1$		\\
			$E${\tiny $(A,B)$}		&	$2$		&	$0$	&	$2$	&	$-1$	&	$0$		\\
			$T_1${\tiny $(A,E_+,E_-)$}	&	$3$		&	$1$	&	$-1$	&	$0$	&	$-1$		\\
			$T_2${\tiny $(B,E_+,E_-)$}	&	$3$		&	$-1$	&	$-1$	&	$0$	&	$1$		\\
			\hline
		\end{tabular}
	\end{minipage}\hfill
	\begin{minipage}[t]{0.4\textwidth}
		\vspace{-1.46cm}
		\begin{tabular}{|c|rrrr|r|}
			\hline
			$C_4$	&	$\mathbb{1}$	&	$C_{4_1}$	&	$C_2$	&	$C_{4_3}$	&	TR	\\
			\hline
			$A$	&	$1$		&	$1$		&	$1$	&	$1$		&	(a)	\\
			$B$	&	$1$		&	$-1$		&	$1$	&	$-1$		&	(a)	\\
			$E_+$	&	$1$		&	$i$		&	$-1$	&	$-i$		&	(b)	\\
			$E_-$	&	$1$		&	$-i$		&	$-1$	&	$i$		&	(b)	\\
			\hline
		\end{tabular}
	\end{minipage}
\end{minipage}

	\caption{Character tables for the $O$ and $C_4$ point groups relevant for the $H$ ($\Gamma$) point and $\Delta$ line (Fig.~\ref{fig:8srs-cross-section} right), respectively. In the conventional manner, the rows cover the irreducible representations and the columns the symmetry operations. The $C_4$ representations that are included in each $O$ representation (the compatibility relations) are listed in brackets on the left. The time reversal symmetry type TR for the $C_4$ group is further added in the last column.}
	\label{tab:O-C_4}
\end{table}

Despite the absence of CD and ellipticity below $\Omega_c$, the rotation angle goes up to $\approx-8^\circ$ at $\Omega\approx0.6$ even within the fundamental bands. This non-optimized PC exhibits therefore roughly $1/3$ of the optical rotation that can be achieved with metallic metamaterials \cite{DeckerZhaoSoukoulisLindenWegener:2010} operating at comparable wavelengths in the near-infrared. Transmission is almost $1$ at those wavelengths and dominated by a single mode process so that an effective medium approach is justified. The {\bf 8-srs} operating in the upper fundamental band frequency region is hence a promising candidate for an optically active and lossless metamaterial.

In conclusion, the {\bf 8-srs} is a prototype for a lossless chiral PC material that provides strong optical rotary power combined with zero ellipticity below a frequency threshold that is way above the fundamental band edges for reasonable dielectric contrast. This unique combination of desired chiro-optical behaviour makes it a good candidate for optical rotators or circular polarization beam splitters \cite{TurnerSabaZhangCummingSchroederTurkGu:2013}. Further, the scattering process with a PC slab is highly non-linear so that optical activity changes rapidly at the frequencies where poles in the generalized Airy formula exist \cite{ByrneBottonAsatryanNicoroviciNortonMcPhedranSterke:2006}. It therefore also is an attractive metamaterial that could be used for optical switches which is fully scalable in contrast to e.g.~liquid crystals in a smectic nematic phase.

Note that {\bf 1-srs} and {\bf 2-srs} PCs have been manufactured on the micron-scale with $a_0=1.2\mu m$ by direct laser writing (DLW) methods \cite{TurnerSchroederTurkGu:2011,TurnerSabaZhangCummingSchroederTurkGu:2013}. The DLW fabrication of the {\bf 8-srs} appears therefore feasible at a structure size such that it exhibits all the basic features predicted by our theory in the near infrared.

While the {\bf 8-srs} geometry is a particularly interesting design from a chiral-optical perspective, it is important to note that results (a-f) hold for {\em any} PC structure with symmetry $I432$. Even more generally, rules for the scattering matrix (d-f) are valid for any chiral PC with a $4$-fold rotational symmetry in propagation direction.

Finally, we have restricted our representation analysis to the $H$ point of lowest vacuum frequency $H^{(0)}$ for the sake of simplicity. The reduction procedure, however, is equivalent for all higher $H$ or $\Gamma$ points with $O$ symmetry yielding for example the irreducible representations $2A_2+2E+4T_1+2T_2$ at $\Gamma^{(1)}$ and $2E+2T_1+2T_2$ at $H^{(1)}$. Furthermore, other directions or symmetries can be treated in the same way. Preliminary investigation for $\vec{k}$ along $[111]$ (a $3$-fold axis in general) yields the same main features as for $\vec{k}$ on $\Delta$.

\begin{acknowledgments}
We thank Stephen Hyde for the inspiration to analyse multiple inter-threaded network structures, and Nadav Gutman for pointing out the potential of group theory. We thank Michael Fischer for comments on the manuscript. MS, KM and GST acknowledge funding by the German Science Foundation through the Cluster of Excellence {\em Engineering of Advanced Materials}. MT and MG acknowledge funding by the Australian Research Council through the {\em Centre for Ultrahigh-Bandwidth Devices for Optical Systems} (project CE110001018). 
\end{acknowledgments}

\bibliography{8srs-literature}

\end{document}